\documentclass{scrartcl}
\usepackage[utf8]{inputenc}

\title{Measuring the Speed of Sound in Air Using a Smartphone and a Cardboard Tube}
\author{Simen Hellesund \\ \large{University of Oslo}}
\date{}

\usepackage{graphicx}
\usepackage{multicol}
\usepackage[T1]{fontenc}
\usepackage{float}
\usepackage{amsmath}
\usepackage{upgreek} 
\usepackage{dcolumn}
\usepackage{booktabs}
\usepackage{url}
\usepackage{cite} 
\usepackage{textcomp} 

\begin{document}\sloppy

\twocolumn[{%
\maketitle
\section*{Abstract}
This paper demonstrates a variation on the classic Kundt's tube experiment for measuring the speed of sound. The speed of sound in air is measured using a smartphone and a cardboard tube, making the experiment very economical in terms of equipment. The speed of sound in air is measured to within 3\% of the theoretical prediction. \vspace{.3em}

\textit{Keywords}: Speed of sound, Smartphone, Kundt's tube.

\vspace{5mm}
}]

\section*{Introduction}

In recent years, smartphones have become packed with sensors; microphones, cameras, accelerometers, magnetometers, gyroscopes, thermometers, proximity sensors etc. Having become ubiquitous in society, smartphones can provide economical alternatives to expensive laboratory equipment in physics education.

Several papers have examined the use of smartphones in acoustic experiments for educational purposes\cite{cit:Kuhn,cit:Kuhn2,cit:Parolin2,cit:Klein,cit:Gomez}. A common experiment in physics education is to measure the speed of sound \textit{c} in air, or other gasses, by observing standing acoustic waves in a tube. August Kundt first described this experiment in 1866\cite{cit:Kundt}. Such an experiment is therefore often referred to as \textit{Kundt's tube}.

Parolin and Pezzi have shown how the experiment can be performed using two smartphones\cite{cit:Parolin}. Yavuz has shown how it can be done using a single smartphone by partially submerging the tube in water\cite{cit:Yavuz}. The aim of this paper is to attempt to perform the experiment using only a smartphone and a cardboard tube, offering an alternative method to the one outlined in \cite{cit:Yavuz}.

\section*{Theory}
For a sinusoidal wave with constant frequency \textit{f} and wavelength $\lambda$, propagating in a medium, the speed of sound in said medium is given by:
\begin{equation}
    c = \lambda f\,.
\label{eq:c}
\end{equation}
This means that if we can determine both the frequency and wavelength of the wave, we can measure the speed of sound in the medium. For this experiment, the medium in question is air at room temperature and atmospheric pressure.

When an acoustic wave enters through the open end of a half-closed tube and hits the closed end, part of the wave is reflected back down the tube towards the open end. At specific wavelengths, the incident and the reflected wave form a standing wave. In the \textit{antinodes} of the standing wave, the points on the standing wave where the amplitude is maximal, the amplitude of the standing wave is greater than the amplitude of the incident wave alone. The opening of the tube will always be a displacement antinode of the standing waves. The wavelengths at which the standing waves occur are called the \textit{resonance wavelengths} of the tube. For the half-closed tube, the resonances occur when the length of the tube equals an odd number of quarter wavelengths of the incident wave:
\begin{equation}
\lambda_n = \frac{4L}{n}, \qquad n=1,3,5,...\,.
\label{eq:harmonic}
\end{equation}
The number \textit{n} is often referred to as the \textit{n-th harmonic} of the tube. \textit{L} is the length of the tube.

The \textit{resonance frequencies} $f_n$ of the tube, the frequencies at which standing waves occur in the tube, can be found by combining equations \ref{eq:c} and \ref{eq:harmonic}:
\begin{equation}
    f_n = \frac{cn}{4L}, \qquad n=1,3,5,... \,.
\end{equation}
We see that the resonance frequencies as a function of \textit{n} is the equation of a straight line. The slope \textit{a} of this line is given by
\begin{equation}
    a = \frac{c}{4L}\,.
\end{equation}
This means that if we can identify the resonance frequencies of the tube and fit a straight line to them, we can calculate the speed of sound as
\begin{equation}
    c = 4aL\,.
    \label{eq:cMeas}
\end{equation}

It turns out that the \textit{acoustic length} of the tube is slightly longer than its physical length. The position of the antinode at the tube's open end will be a small distance outside of the tube. A more accurate measurement can therefore be performed by adding a correction term $\delta L$ to the length of the tube. Levine and Schwinger found this correction term to be $\delta L = 0.61r$, where $r$ is the radius of the tube\cite{endCorrection}. Using this correction to the length of the tube, the speed of sound is given by
\begin{equation}
    c = 4a(L + 0.61r)\,.
    \label{eq:cMeasCorr}
\end{equation}

The theoretical speed of sound in air $c_{\mathrm{T}}$ can be calculated as
\begin{equation}
    c_{\mathrm{T}} = \sqrt{\frac{\gamma R T}{M}}\,,
    \label{eq:cTheory}
\end{equation}
where $\gamma$ is the \textit{adiabatic index} of air; \textit{R} is the \textit{molar gas constant}; \textit{T} is the temperature of the air in Kelvin; and \textit{M} is its \textit{molar mass}. Equation \ref{eq:cTheory} is only valid for an ideal gas. At room temperature and normal atmospheric pressure, air behaves close enough to an ideal gas for our purpose.

\section*{Setup}
The following equipment is used for this experiment: a smartphone with a signal generator app and a recording app installed; a cardboard tube, closed in one end; a thermometer; a tape measure; a computer.

The smartphone used is a Motorola Moto g6 Plus. There are several apps available both for generating audio sine waves and for recording. The function generator app must be able to perform a sweep over frequencies. The generator app used in this experiment is called \textit{Function Generator} and is available in the Google Play Store\cite{cit:Signal}. The recording app must record audio with a sampling frequency known to the user. The recording app used for this experiment is called  \textit{Smart Recorder}, also available in the Google Play Store\cite{cit:Recoding}. For users of Apple products, the recording app \textit{Voice Recorder Lite: HD Audio Recording \& Playback}\cite{cit:RecodingApple} and the signal generator \textit{Audio Function Generator}\cite{cit:SignalApple}, both available from the Apple App Store, may be used.

\begin{figure}[ht]
    \centering
    \includegraphics[width=.5\textwidth]{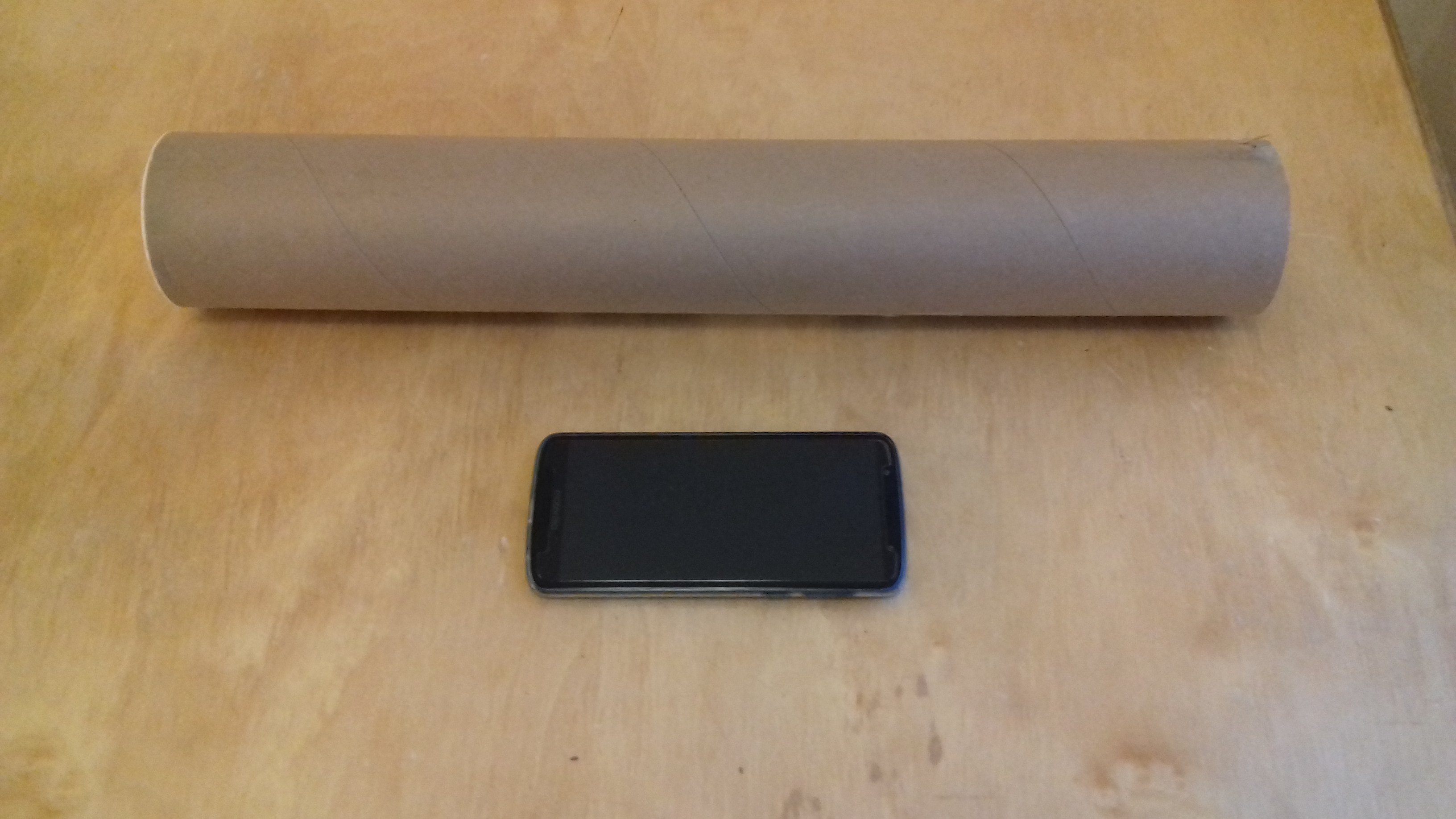}
    \caption{The smartphone and the cardboard tube used in the experiment.}
    \label{fig:Setup}
\end{figure}

The cardboard tube used is one intended for storing or shipping posters. One end is stopped by a plastic plug. The length of the tube is measured, using the tape measure, to be 47.6 cm. The inner diameter of the tube is measured to be 7.5 cm. The tube and the smartphone used in the experiment can be seen in Figure \ref{fig:Setup}.


The experiment is performed in an anechoic chamber at the physics department at the University of Oslo. Having access to such a room is not critical to the experiment, although one should aim to limit background noise as much as possible.

The thermometer is not used in the measurement directly but is used to measure the temperature in the room during the experiment. This is then used to calculate $c_{\mathrm{T}}$. The temperature in the anechoic chamber is measured to be 24 \textdegree C. 

\section*{Procedure}
\begin{figure}[ht]
    \centering
    \includegraphics[width=.5\textwidth]{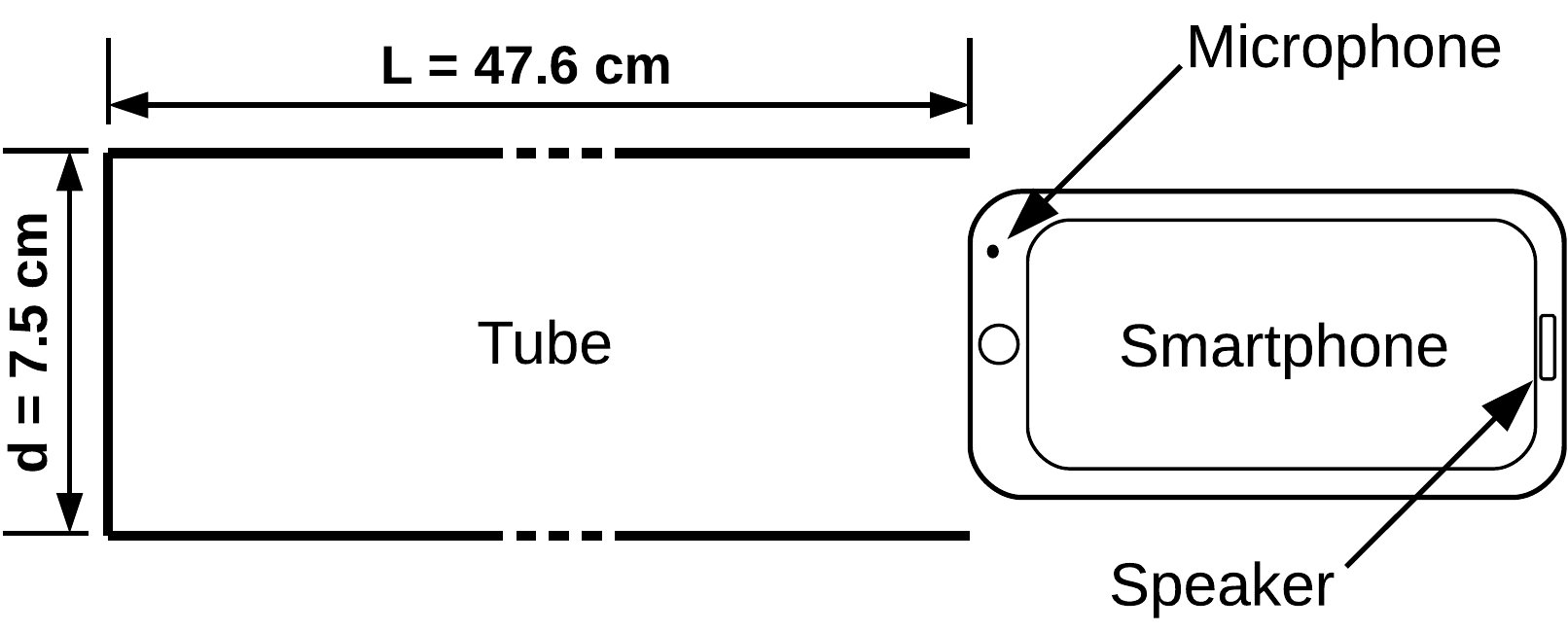}
    \caption{Schematic illustrating the placement of the tube and smartphone during the experiment.}
    \label{fig:schematic}
\end{figure}

We place the smartphone such that the microphone is located in the opening of the tube. This is shown in Figure \ref{fig:schematic}. The phone is set to record audio with a sampling frequency of 44.1 kHz. 
While the phone is recording, the function generator app emits a pure sine wave. The sine wave sweeps from 50 Hz to 3000 Hz at a rate of 1 Hz/s. The audio recording is stored in .wav format. This format makes for easy data analysis later.


\section*{Measuring Frequency}
We need to know the frequencies at which the resonances occur. The function generator does not store any information on which frequency is emitted at which time. Thus, the frequency has to be inferred from the recording.

Frequency cannot be instantaneously measured. The signal must be monitored over some period of time to count how many times it oscillates per second.

\begin{figure}[ht]
    \centering
    \includegraphics[width=.5\textwidth]{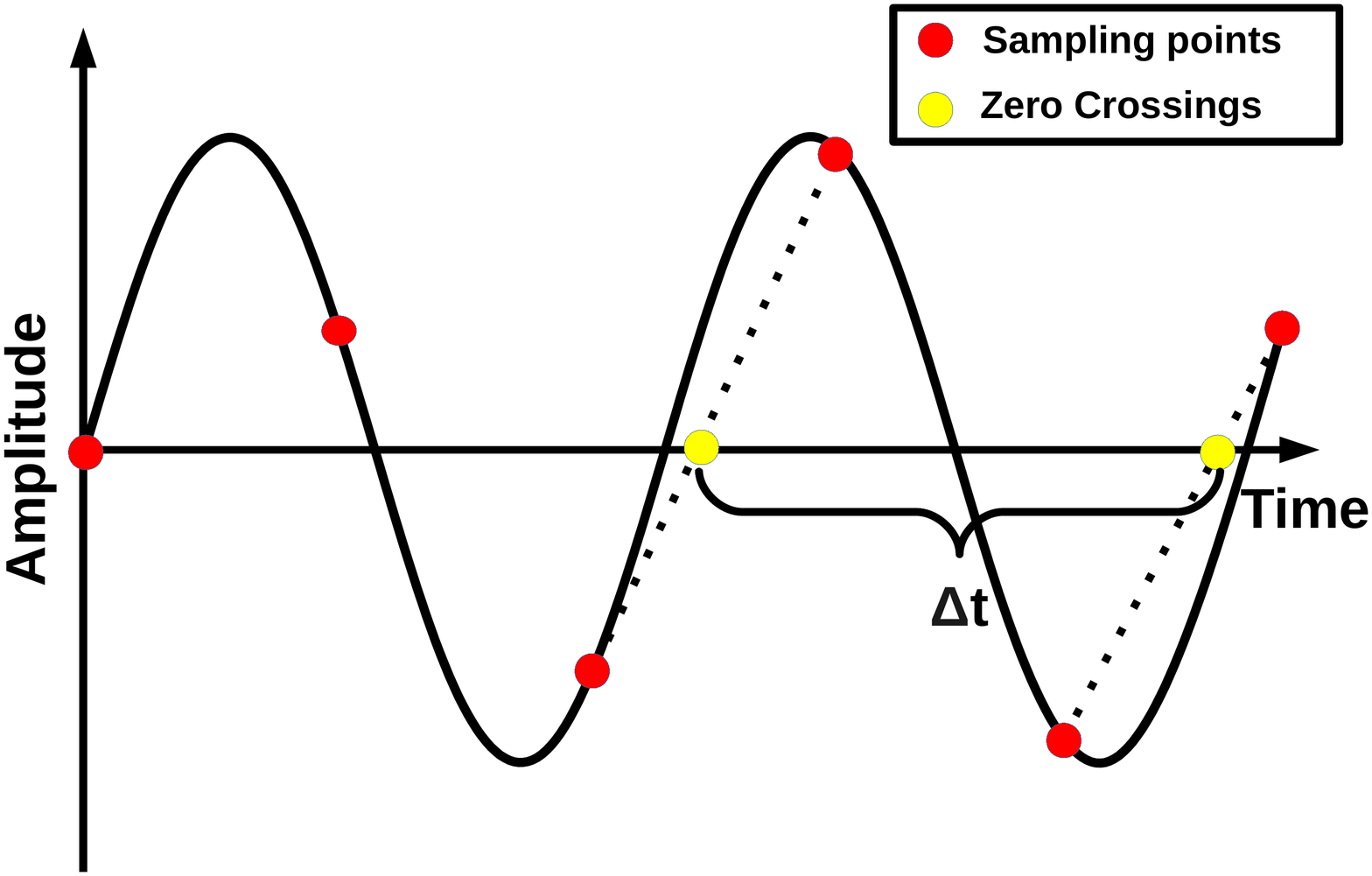}
    \caption{The Zero Crossing Method Illustrated. The black line is the pure sine wave signal. The red points are the points sampled by the smartphone. The yellow points are the zero crossing points used to approximate the frequency of the signal.}
    \label{fig:ZCM}
\end{figure}

In this case, where we are trying to identify a pure sine wave in a high signal to noise ratio environment, we can use the method known as the \textit{Zero Crossing Method} (ZCM)\cite{cit:ZCM}. The ZCM works by determining the points in time where the waveform of the recording crosses from negative to positive values (or vice versa). These points are approximated by identifying the points on either side of such a crossing and drawing a straight line between them. The zero crossing point of the signal is approximated as the point at which this straight line becomes zero. The distance from one such zero crossing point to the next is an approximation of the period of oscillation of the signal. The inverse of this period is the frequency of the signal. The ZCM procedure is illustrated in Figure \ref{fig:ZCM}.

\begin{figure}[ht]
    \centering
    \includegraphics[width=.5\textwidth]{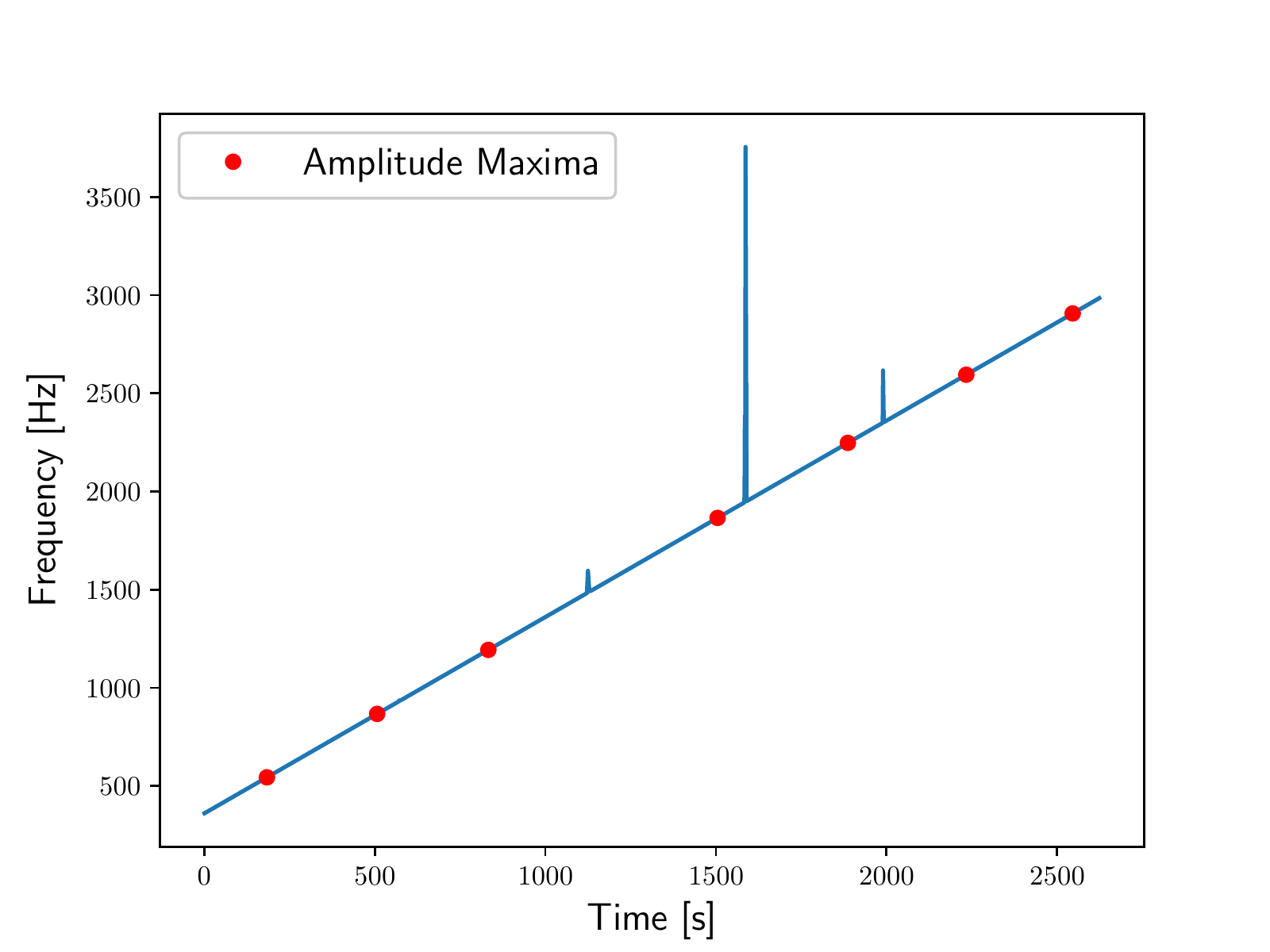}
    \caption{Frequencies measured using the Zero Crossing Method.}
    \label{fig:frequencies}
\end{figure}

The samples of the audio recording are split into one-second intervals. In each of these intervals, the ZCM is used to find all the zero crossings. The measured frequency in each interval is taken to be the mean of all the frequencies measured from these zero crossings. These frequencies are plotted as a function of time in Figure \ref{fig:frequencies}.

Looking at Figure \ref{fig:frequencies}, we notice that there are a few points where the ZCM clearly fails to identify the frequency emitted by the function generator. In our case, these points can be ignored, as they are far away from the points where the sound intensity of the recording peaks.

The results obtained using the ZCM are cross-checked by measuring the frequency in the recording using a \textit{Fast Fourier Transfer} (FFT) in each second interval. The most prominent frequency in the FFT spectrogram of each interval is taken to be the frequency of the signal. This method yields identical results to those obtained by the ZCM. However, the FFT method is significantly slower and is therefore not used in the final analysis.

\section*{Data Analysis}
The start of the audio recording proves too noisy to be of use. The low frequencies emitted by the function generator are perhaps causing the smartphone to vibrate. This region of the recording is therefore cut before the analysis.

The data analysis is performed in Python. The SciPy python library contains the packages needed for reading the .wav file, as well as for signal processing and curve fitting\cite{cit:SciPy}.

\begin{figure}[ht]
    \centering
    \includegraphics[width=.5\textwidth]{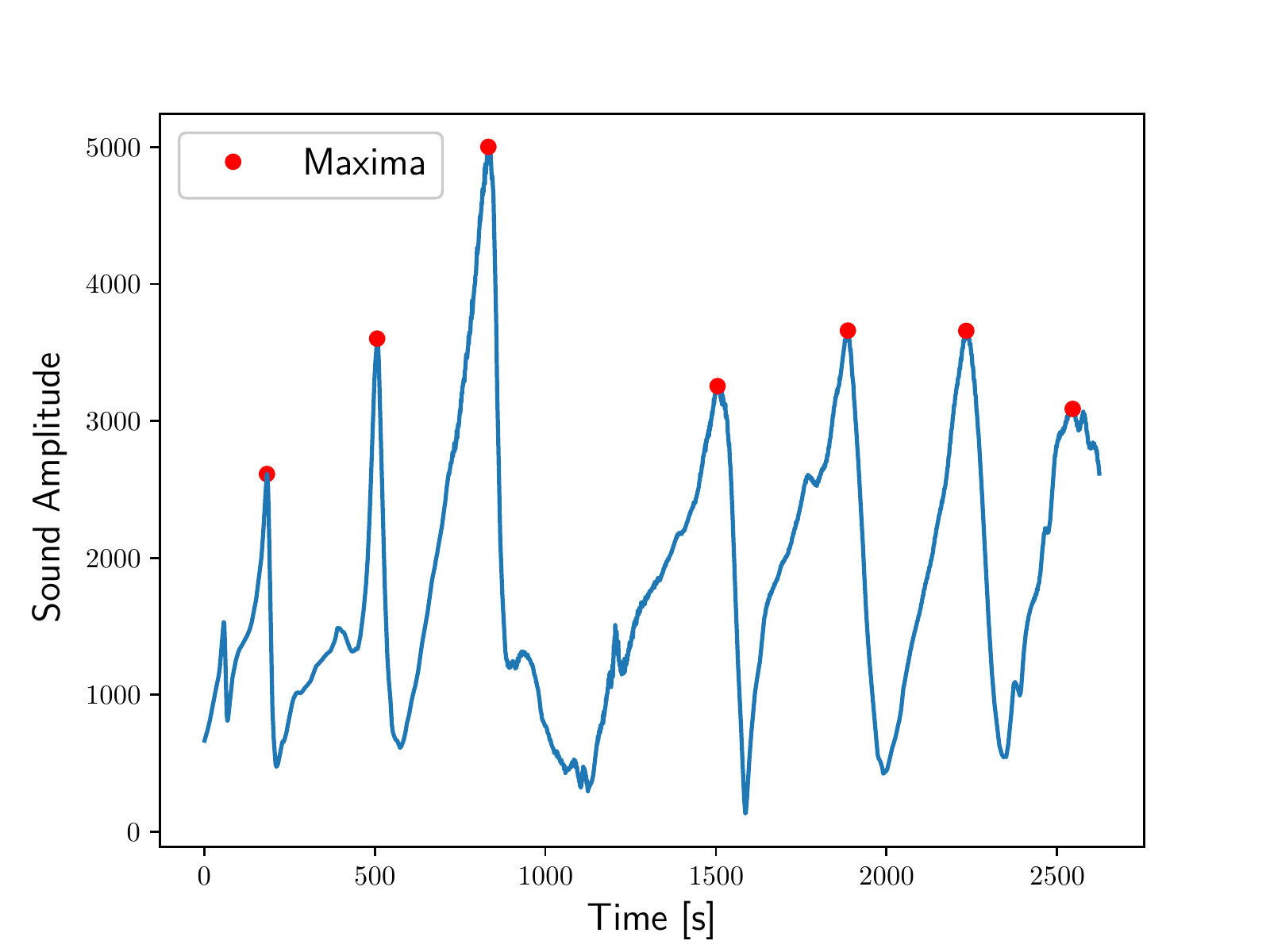}
    \caption{Sound amplitude of the audio recording.}
    \label{fig:intensity}
\end{figure}

We need to identify the points where the sound intensity of the recording is maximal, as these maxima will occur at the resonance frequencies of the tube. First, the recording is split into the same one-second increments used to measure the frequency. The absolute value of the digital samples is taken in each of these intervals. The peaks in the resulting signal are identified using the \texttt{find\_peaks} function of the SciPy signal processing package. The mean height of the peaks found in each second increment is taken to be the sound amplitude. The resulting distribution is shown in Figure \ref{fig:intensity}.

Next, the \texttt{find\_peaks} function is used once more to identify the peaks in the sound amplitude distribution. These peaks are also shown in Figure \ref{fig:intensity}. The frequencies measured at the times of these peaks are the resonance frequencies of the cardboard tube. These frequencies are recorded in Table \ref{tab:Frequencies} in the results section. It seems one resonance frequency fails to give a peak in the sound amplitude distribution. The reason for this is unknown. The frequencies at $n=7$ and above are therefore shifted up by one.

\section*{Results}
\begin{figure}[ht]
    \centering
    \includegraphics[width=.5\textwidth]{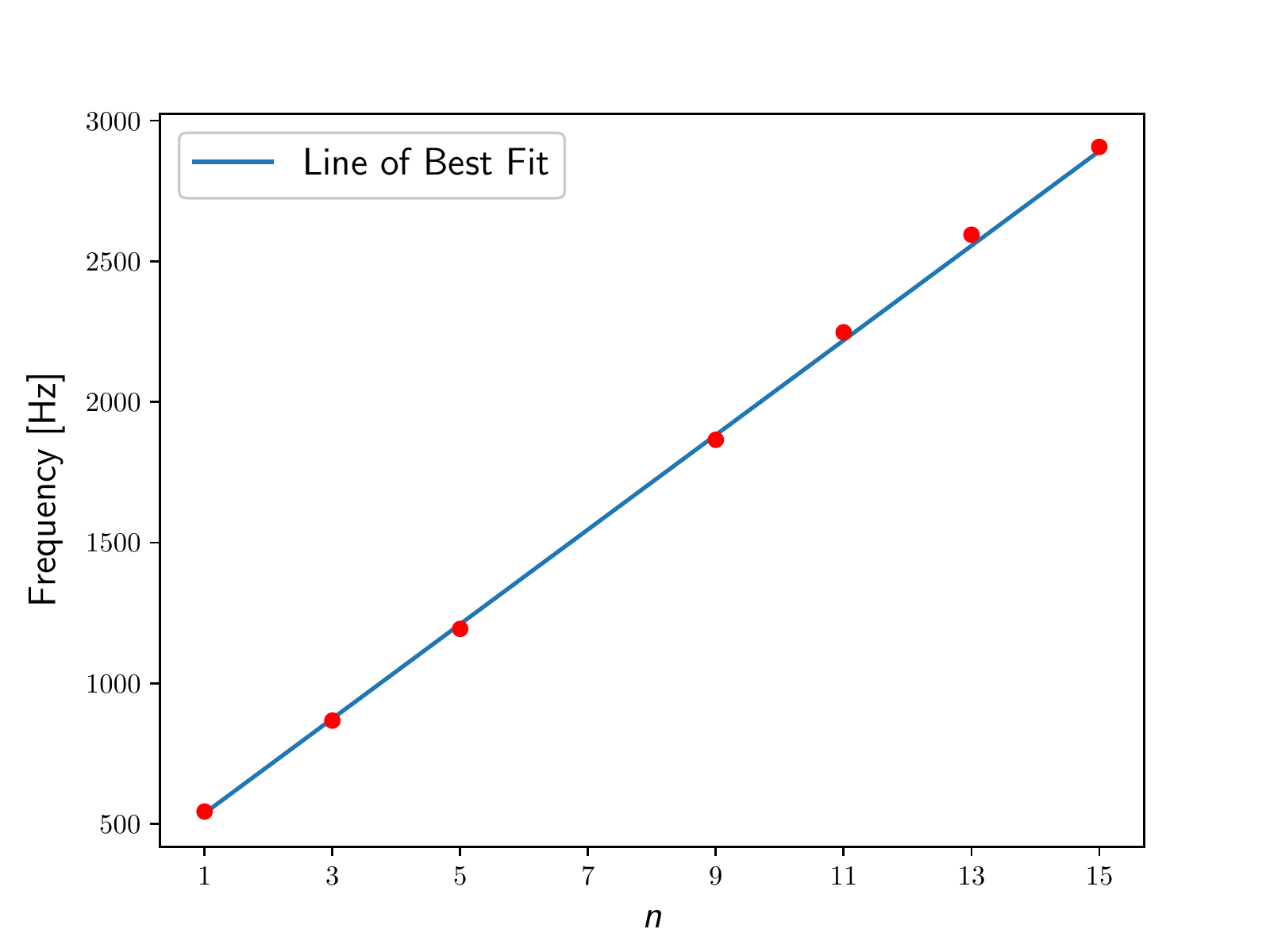}
    \caption{Resonance frequencies of the cardboard tube with line of best fit.}
    \label{fig:Fit}
\end{figure}

The resonance frequencies identified in the previous section are listen in Table \ref{tab:Frequencies}. A straight line is fitted to these frequencies as a function of \textit{n} using the \texttt{polyfit} function of the Python module NumPy\cite{cit:numpy}. The frequencies as well as the line of best fit is shown in Figure \ref{fig:Fit}. The slope of the straight line is used to calculate \textit{c} using Equation \ref{eq:cMeasCorr}. The speed of sound in air at 24 \textdegree C is measured to be $c=335$ m/s. 



Using Equation \ref{eq:cTheory}, the theoretical speed of sound in an ideal gas at 24 \textdegree C is calculated to be $c_{\mathrm{T}}=345$ m/s. The measured value of \textit{c} is thus within 3\% of the theoretical prediction.

\begin{table}[ht]
\caption{Frequencies at peak amplitudes.}
\label{tab:Frequencies}
\centering
\begin{tabular}{lc}
\toprule
n     &
\multicolumn{1}{c}{Frequency [Hz]} \\ \hline
1  & 544                \\
3  & 867             \\
5  & 1130               \\
7  & Unknown                 \\
9  & 1865              \\
11 & 2247              \\
13 & 2594               \\
15 & 2906           \\
\bottomrule
\end{tabular}
\end{table}

\section*{Conclusion}
In this paper, a method for measuring the speed of sound in air has been demonstrated using a smartphone and a cardboard tube. The experiment is very economical, and can thus be performed in places with limited access to laboratory equipment.

The experiment requires some amount of programming. It can provide learning opportunities for students both in experimental methods as well as in data analysis.

\bibliographystyle{unsrt}
\bibliography{biblio.bib}

\end{document}